# Teleparallel Complex Gravity as Foundation for Noncommutative Gravity

Hitoshi Nishino[1] and Subhash Rajpoot[2]

*Department of Physics & Astronomy*
*California State University*
*Long Beach, CA 90840*

## Abstract

We present a teleparallel complex gravity as the foundation for the formulation of noncommutative gravity theory. The negative energy ghosts in the conventional formulation with $U(1,3)$ local Lorentz connection no longer exists, since the local Lorentz invariance is broken down to $U(1,3)$ global Lorentz symmetry. As desired, our teleparallel complex gravity theory also passes the key classical test of perihelion advance of Mercury. Based on this result, we present a lagrangian for the noncommutative teleparallel gravity theory.

---

[1] E-Mail: hnishino@csulb.edu
[2] E-Mail: rajpoot@csulb.edu

## 1. Introduction

The recent developments of open strings or D-branes lead to the consideration of constant background antisymmetric field $B_{\mu\nu}$, which in turn implies that the coordinates of space-time should be noncommutative [1]. It has been well-known that the noncommutative generalization [1] of gravity theory necessarily needs a complex metric [2], because the introduction of the $\star$ product with $i\theta^{\mu\nu}$ makes the metric complex. However, once the metric becomes complex, we must consider new components present in the theory, such as the antisymmetric component $B_{\mu\nu} \equiv Im(g_{\mu\nu})$. In particular, the presence of its kinetic term, or the antisymmetric part of the vierbein gravitational field equation $\partial S/\partial e_\mu{}^a = 0$ of the total action $S$ should be studied in the light of a consistent gravitational theory.

As the first attempt to formulate such a complex gravity as the preliminary for noncommutative gravity, a lagrangian has been presented in [2]. However, the problem in this formulation was that the new components $B_{\mu\nu}$ acquire a kinetic term, and some of its components become non-physical [3]. These components are negative energy ghosts, and are not acceptable at the level of classical field theory.

The origin of such ghost components can be traced back to the introduction of $U(1,3)$ local Lorentz symmetry in the system, $i.e.$, the introduction of the Lorentz connection $\omega_{\mu a}{}^b$ as its gauge field. A similar situation has been encountered in noncommutative non-Abelian gauge field theories. This is because the ordinary noncommutative gauge theories require that the gauge groups to be $U(n)$. To avoid this problem, certain formulation that enables the gauge groups to be other than $U(n)$, such as $SO(n)$ or $Sp(n)$, has been presented [4]. Applying similar techniques to this $U(1,3)$ local Lorentz symmetry, alternative noncommutative gravity theories were formulated based on the noncommutative diffeomorphism $ISO(1,3)$ group [5], or $CSO(3,1)$ group for complex symmetric metric [6].

In this Letter, we present a different approach as a remedy for the negative energy ghosts [3], by freezing the $U(1,3)$ local Lorentz symmetry into to a global one. In ordinary gravity theory with real metrics alone, such a formulation is sometimes called 'teleparallel gravity formulation' [7] in which the $SO(1,3)$ local Lorentz symmetry is frozen down to a global $SO(1,3)$ symmetry, and therefore there is no gauge field or spin connection $\omega_{\mu m}{}^n$ for the Lorentz symmetry. Our strategy for complex gravity is similar, namely, we freeze the $U(1,3)$ local symmetry into a global $U(1,3)$ symmetry, requiring teleparallelism without any introduction of its gauge fields, and thus avoid the problem of negative energy ghosts. The importance of teleparallel gravity in the context of noncommutative geometry has been pointed out in [8], in which teleparallel gravity is shown to arise out of dimensional reduction of noncommutative gauge theory. However, we will rely on a teleparallel gravity theory as the foundation of complex gravity from the outset, in order to resolve the problem with the negative energy ghosts in the $U(1,3)$ local Lorentz covariant formulation [2]. We stress that teleparallel gravity as the foundation of noncommutative gravity is the legitimate starting point, since the constant $\theta^{\mu\nu}$ in noncommutative gravity manifestly breaks Lorentz symmetry.

## 2. Teleparallel Complex Gravity

Since we are considering the explicit breaking of local Lorentz symmetry, it is crucial to understand the degrees of freedom of the vierbein components $e_\mu{}^a$ and its hermitian conjugates $e_{\mu a} \equiv (e_\mu{}^a)^\dagger$.[3] In the local Lorentz covariant formulation [2], all together there are originally $4 \times (4+4) = 32$ components in $e_\mu{}^a$ and $e_{\mu a}$. However, the $U(1,3)$ local Lorentz symmetry with 16 parameters deletes 16 components, leaving only 16 components. These 16 components are equivalent to the symmetric part $G_{\mu\nu} \equiv g_{(\mu\nu)} \equiv Re(g_{\mu\nu})$, and the antisymmetric part $B_{\mu\nu} \equiv -ig_{[\mu\nu]} \equiv Im(g_{\mu\nu})$ of the metric tensor $g_{\mu\nu}$ [2]. In our formulation, on the other hand, the original 32 components are not deleted by the $U(1,3)$ local Lorentz symmetry, and all of them are intact. This formulation has the advantage of deleting the kinetic term for the $B$-field. The price to be paid is that there are 16 additional components in the vierbeins whose effect must be carefully investigated.

We first give preliminaries for the formulation for teleparallel complex gravity. If the $U(1,3)$ local Lorentz symmetry is manifest from the outset, we must introduce the Lorentz connection for local Lorentz covariance. This causes the problem of negative energy ghosts [2] which we would like to avoid. Therefore, it is natural to consider the formulation in which local Lorentz symmetry is not built-in, or at least it is not manifest from the outset.

The most fundamental relationships among geometric quantities, such as the vierbein and metric are

$$
\begin{aligned}
&(\eta_a{}^b) = \text{diag.}(-,+,+,+) \ , \\
&e_\mu{}^a e_a{}^\nu \equiv \delta_\mu{}^\nu \ , \quad e_a{}^\mu e_\mu{}^b \equiv \delta_a{}^b \ , \quad e_{\mu a} \equiv (e_\mu{}^a)^\dagger \ , \quad e^{a\mu} \equiv (e_a{}^\mu)^\dagger \ , \\
&(g_{\mu\nu})^\dagger \equiv g_{\nu\mu} \ , \quad (g^{\mu\nu})^\dagger \equiv g^{\nu\mu} \ , \quad e \equiv \det(e_\mu{}^a) \ , \quad \overline{e} \equiv e^\dagger \equiv \det(e_{\mu a}) \ , \\
&g_{\mu\nu} \equiv e_{\mu a}\eta_b{}^a e_\nu{}^b \ , \quad g^{\mu\nu} \equiv e_a{}^\mu \eta_b{}^a e^{b\nu} \ , \quad g_{\mu\nu}g^{\nu\rho} = \delta_\mu{}^\rho \ , \quad g^{\mu\nu}g_{\nu\rho} = \delta_\rho{}^\mu \ ,
\end{aligned} \quad (2.1)
$$

where the symbol $\dagger$ is for hermitian conjugations. As usual, the metric $g_{\mu\nu}$ has both symmetric and antisymmetric components. The most basic global $U(1,3)$ transformation rules are

$$
\begin{aligned}
&\delta_\alpha e_\mu{}^a = -\alpha_b{}^a e_\mu{}^b \ , \quad \delta_\alpha e_a{}^\mu = +\alpha_a{}^b e_b{}^\mu \ , \\
&\delta_\alpha e_{\mu a} = +\eta_a{}^c \alpha_c{}^d \eta_d{}^b e_{\mu b} \ , \quad \delta_\alpha e^{a\mu} = -\eta_c{}^a \alpha_d{}^c \eta_b{}^d e^{b\mu} \ , \\
&(\alpha_a{}^b)^\dagger = -\eta_b{}^c \alpha_c{}^d \eta_d{}^a \ ,
\end{aligned} \quad (2.2)
$$

where $\alpha_a{}^b$ is the space-time independent parameters for our global $U(1,3)$, complying with the notation in [2]. Accordingly, the metric itself does not transform: $\delta_\alpha g_{\mu\nu} = 0$, $\delta_\alpha g^{\mu\nu} = 0$. Relevantly, the $U(1,3)$ invariant product is $(U_a)^\dagger \eta_a{}^b V_b$, because $\delta_\alpha[(U_a)^\dagger \eta_a{}^b V_b] = 0$. For simplicity, we use the *bars* instead of the *daggers* whenever it is not confusing, such as $\overline{e} \equiv e^\dagger$ in (2.1).

---

[3] We follow the notation of ref. [2] in this paper, unless otherwise noted.

There are other important geometrical equations for later purposes. One of them is the definition of the covariant derivative:

$$D_\mu V^\nu \equiv \partial_\mu V^\nu + \Gamma_{\mu\rho}{}^\nu V^\rho ~, \tag{2.3}$$

for a complex vector $V^\nu \equiv V^a e_a{}^\nu$. Since the vierbein $e_a{}^\nu$ but *not* its hermitian conjugate $e^{a\nu}$ is used here, we use $D_\mu$ instead of its hermitian conjugate $\overline{D}_\mu$. The latter is used, when we take the hermitian conjugate of the whole equation of (2.3):

$$\overline{D}_\mu \overline{V}^\nu \equiv \partial_\mu \overline{V}^\nu + \overline{\Gamma}_{\mu\rho}{}^\nu \overline{V}^\rho ~, \tag{2.4}$$

for $\overline{V}^\nu \equiv \overline{V}_a e^{a\nu}$. Relevantly, the commutation relations between the $D_\mu$'s and the resulting Bianchi identity are

$$[D_\mu, D_\nu] = -C_{\mu\nu}{}^\rho D_\rho ~, \qquad D_{[\mu} C_{\nu\rho]}{}^\sigma + C_{[\mu\nu|}{}^\tau C_{\tau|\rho]}{}^\sigma \equiv 0 ~, \tag{2.5a}$$

$$[\overline{D}_\mu, \overline{D}_\nu] = -\overline{C}_{\mu\nu}{}^\rho \overline{D}_\rho ~, \qquad \overline{D}_{[\mu} \overline{C}_{\nu\rho]}{}^\sigma + \overline{C}_{[\mu\nu|}{}^\tau \overline{C}_{\tau|\rho]}{}^\sigma \equiv 0 ~, \tag{2.5b}$$

where $C_{\mu\nu}{}^a$ are anholonomy coefficients:

$$C_{\mu\nu}{}^\rho \equiv C_{\mu\nu}{}^a e_a{}^\rho ~, \qquad C_{\mu\nu}{}^a \equiv \partial_\mu e_\nu{}^a - \partial_\nu e_\mu{}^a ~,$$

$$\overline{C}_{\mu\nu}{}^\rho \equiv (C_{\mu\nu}{}^\rho)^\dagger = \overline{C}_{\mu\nu\,a} e^{a\rho} ~, \qquad \overline{C}_{\mu\nu\,a} \equiv (C_{\mu\nu}{}^a)^\dagger = \partial_\mu e_{\nu a} - \partial_\nu e_{\mu a} ~,$$

$$\overline{C}^{\mu\nu}{}_\rho \equiv (C^{\mu\nu}{}_\rho)^\dagger = (g^{\sigma\mu} g^{\tau\nu} g_{\rho\lambda} C_{\sigma\tau}{}^\lambda)^\dagger = g^{\mu\sigma} g^{\nu\tau} g_{\lambda\rho} \overline{C}_{\sigma\tau}{}^\lambda ~, \qquad etc. \tag{2.6}$$

The $D_\mu$'s in (2.5a) uses only $\Gamma_{\mu\nu}{}^\rho$ but none of its hermitian conjugates $\overline{\Gamma}_{\mu\nu}{}^\rho$, because $C_{\mu\nu}{}^\rho$ is composed only of $e_\mu{}^a$ and $e_a{}^\mu$ but none of their hermitian conjugates $e_{\mu a}$ and $e^{a\mu}$, as seen from (2.6). If we had manifest local Lorentz covariance, there would be an additional term in (2.5b) proportional to the Lorentz curvature tensor. This term is now absent, due to the lack of manifest Lorentz covariance in our teleparallel gravity.

Another important equation comes from the vierbein postulate that leads to the expression of $\Gamma_{\mu\nu}{}^\rho$ in terms of vierbein:

$$D_\mu e_\nu{}^a = \partial_\mu e_\nu{}^a - \Gamma_{\mu\nu}{}^\rho e_\rho{}^a = 0 \quad \Longrightarrow \quad \Gamma_{\mu\nu}{}^\rho = e_a{}^\rho \partial_\mu e_\nu{}^a ~, \qquad \overline{\Gamma}_{\mu\nu}{}^\rho = e^{a\rho} \partial_\mu e_{\nu a} ~. \tag{2.7}$$

The reason for using $D_\mu$'s instead of $\overline{D}_\mu$ in the first equation here has been already stated, and shows how important it is to distinguish $\Gamma_{\mu\nu}{}^\rho$ from its hermitian conjugate $\overline{\Gamma}_{\mu\nu}{}^\rho$.

## 3. Teleparallel Complex Gravity – Lagrangian and Field Equations

Once the transformation properties of basic quantities are in place, we are ready to present a lagrangian which yields the usual Einstein-Hilbert action, with no manifest $U(1,3)$ local Lorentz covariance.

It is worth noting that we have to be cautious about the significance of local Lorentz symmetry. In the ordinary case with real metric with the familiar $SO(1,3)$ local Lorentz symmetry, even if we start with quantities such as the anholonomy coefficients $C_{\mu\nu}{}^a$, and write down the lagrangian in terms of its quadratic products, the resulting theory may be still locally Lorentz invariant. The reason is that the local Lorentz symmetry is realized as a 'hidden' symmetry at the lagrangian level. For example, it is well-known [7] that a certain combination of quadratic products of $C_{\mu\nu}{}^a \equiv \partial_\mu e_\nu{}^a - \partial_\nu e_\mu{}^a$ yields an action identically equal to a Einstein-Hilbert action up to a total divergence:

$$S_{\rm EH} \equiv \int d^4x\, e \Big[ -\tfrac{1}{8} C_{\mu\nu a} C^{\mu\nu a} + \tfrac{1}{4} C_{\mu\nu\rho} C^{\nu\rho\mu} + \tfrac{1}{2} (C_{\mu\nu}{}^\nu)^2 \Big] \equiv \int d^4x \left( -\tfrac{1}{4} e R \right) \ . \qquad (3.1)$$

Here the vierbeins are the usual real one $(e_\mu{}^a)^\dagger = e_\mu{}^a$, before considering any complex gravity, and $R$ is the usual scalar curvature in terms of the Riemann-Christoffel connection: $\{{}^\rho_{\mu\nu}\}$. In other words, even though each term in (3.1) is not locally Lorentz invariant, the combination of the quadratic products of $C_{\mu\nu}{}^a$ with the appropriate relative coefficients makes the whole expression locally Lorentz invariant.

We now consider complex gravity. Our construction heavily relies on the usage of the anholonomy coefficients and its hermitian conjugates defined in (2.6). Now our candidate lagrangian is the analog of (3.1), and is given by

$$|e|^{-1} \mathcal{L}_0 \equiv a_1 g^{\mu\rho} g^{\nu\sigma} \eta_a{}^b C_{\mu\nu}{}^a \overline{C}_{\rho\sigma b} + a_2 g^{\mu\sigma} e^{b\nu} e_a{}^\rho C_{\mu\nu}{}^a \overline{C}_{\rho\sigma b} + a_3 g^{\mu\rho} e^{b\sigma} e_a{}^\nu C_{\mu\nu}{}^a \overline{C}_{\rho\sigma b} \qquad (3.2a)$$

$$= a_1 g^{\mu\nu} g^{\rho\sigma} g_{\lambda\omega} C_{\mu\rho}{}^\omega \overline{C}_{\nu\sigma}{}^\lambda + a_2 g^{\mu\sigma} C_{\mu\nu}{}^\rho \overline{C}_{\rho\sigma}{}^\nu + a_3 g^{\mu\nu} C_\mu \overline{C}_\nu \qquad (3.2b)$$

$$= a_1 C^{\nu\sigma}{}_\lambda \overline{C}_{\nu\sigma}{}^\lambda + a_2 C^\sigma{}_\nu{}^\rho \overline{C}_{\rho\sigma}{}^\nu + a_3 C^\nu \overline{C}_\nu \ . \qquad (3.2c)$$

Here $|e|^2 \equiv e\, \overline{e}$, while $a_1$, $a_2$ and $a_3$ are real constants. In particular, the case of

$$a_3 = +2 a_2 = -4 a_1 \equiv -4a = +\tfrac{1}{32} \qquad (3.3)$$

is the direct analog of (3.1) in the conventional teleparallel gravity with the real vierbein [7]. The $C_\mu$ and $\overline{C}_\mu$ in (3.2b) are defined by

$$C_\mu \equiv C_{\mu\nu}{}^\nu \ , \quad \overline{C}_\mu \equiv (C_{\mu\nu}{}^\nu)^\dagger = \overline{C}_{\mu\nu}{}^\nu \ , \qquad (3.4)$$

while $C^\sigma{}_\nu{}^\rho$ and $C^\nu$ in (3.2c) are defined by

$$C^\sigma{}_\nu{}^\rho \equiv g^{\tau\sigma} C_{\tau\nu}{}^\rho \ , \quad C^\nu \equiv g^{\rho\nu} C_\rho \ . \qquad (3.5)$$

In terms of these anholonomy coefficients, each term in (3.2c) is manifestly invariant under the global Lorentz transformation (2.2), because

$$\delta_\alpha C_{\mu\nu}{}^\rho = 0 \ , \quad \delta_\alpha \overline{C}_{\mu\nu}{}^\rho = 0 \ . \qquad (3.6)$$

This also explains the reason why we need $\eta_a{}^b$ in the first term in (3.2a), while it is not needed in the second term. To put it differently, in terms of the anholonomy coefficients with curved indices, it is more straightforward to construct globally Lorentz invariant terms for a lagrangian like (3.2c). For readers' convenience, we give the proof of the hermiticity of the second term in (3.2c):

$$(C^\sigma{}_\nu{}^\rho \overline{C}_{\rho\sigma}{}^\nu)^\dagger = (g^{\tau\sigma} C_{\tau\nu}{}^\rho)^\dagger C_{\rho\sigma}{}^\nu = g^{\sigma\tau} \overline{C}_{\tau\nu}{}^\rho C_{\rho\sigma}{}^\nu$$
$$= -g^{\sigma\tau} C_{\sigma\rho}{}^\nu \overline{C}_{\tau\nu}{}^\rho = +C^\tau{}_\rho{}^\nu \overline{C}_{\nu\tau}{}^\rho = C^\sigma{}_\nu{}^\rho \overline{C}_{\rho\sigma}{}^\nu \ , \tag{3.7}$$

despite its 'non-symmetric' appearance.

For explicitness, we give our action after the use of (3.3):

$$S_0 \equiv \int d^4x\, e \Big[ -\tfrac{1}{8} C^{\nu\sigma}{}_\lambda \overline{C}_{\nu\sigma}{}^\lambda + \tfrac{1}{4} C^\sigma{}_\nu{}^\rho \overline{C}_{\rho\sigma}{}^\nu + \tfrac{1}{2} C^\nu \overline{C}_\nu \Big] \equiv \int d^4x\, \mathcal{L}_0 \ . \tag{3.8}$$

Some readers may wonder how the same relative coefficients as in (3.1) can yield a theory now without the kinetic term for $B_{\mu\nu}$. This is because the identity in (3.1) is valid only for a real metric, but we now have the complex metric. As will be seen, our proposed lagrangian (3.8) has no kinetic term for $B_{\mu\nu}$ at the quadratic order, so it is distinct from the lagrangian in [2].

The gravitational field equation from $\mathcal{L}_0$ for the vierbein $e_\mu{}^a$ is

$$F^{\mu\nu} = -2ag^{\rho\mu} D_\sigma \overline{C}^{\sigma\nu}{}_\rho + 2ag^{\rho\mu} D_\sigma \overline{C}_\rho{}^{\sigma\nu} - 2ag^{\rho\mu} D_\sigma \overline{C}_\rho{}^{\nu\sigma} + 4ag^{\nu\mu} D_\rho \overline{C}^\rho - 4ag^{\rho\mu} D_\rho \overline{C}^\nu$$
$$- 2a(C_\rho - iI_\rho)\overline{C}^{\rho\nu\mu} - 4ag^{\sigma\mu}(C_\rho - iI_\rho)\overline{C}_\sigma{}^{[\nu\rho]}$$
$$+ 4ag^{\nu\mu}(C_\rho - iI_\rho)\overline{C}^\rho - 4a(C^\mu - iI^\mu)\overline{C}^\nu$$
$$- 2aC^{\mu\rho}{}_\sigma \overline{C}^\nu{}_\rho{}^\sigma + aC^{\rho\sigma\nu}\overline{C}_{\rho\sigma}{}^\mu + \tfrac{1}{2} ag^{\nu\mu} C^{\rho\sigma}{}_\tau \overline{C}_{\rho\sigma}{}^\tau$$
$$- 2aC^\mu{}_\rho{}^\sigma \overline{C}^\nu{}_\sigma{}^\rho + ag^{\nu\mu} C^\rho{}_\sigma{}^\tau \overline{C}_{\rho\tau}{}^\sigma + 4aC^\mu \overline{C}^\nu - 2ag^{\nu\mu} C^\rho \overline{C}_\rho \doteq 0 \ , \tag{3.9}$$

where

$$F^{\mu\nu} \equiv g^{\rho\mu} F_\rho{}^\nu \equiv g^{\rho\mu} e_\rho{}^a F_a{}^\nu \ , \quad \mathcal{F}_a{}^\mu \equiv |e| F_a{}^\mu \equiv \frac{\delta \mathcal{L}_0}{\delta e_\mu{}^a} \ . \tag{3.10}$$

Here we have already used the condition (3.3) up to $a \equiv -1/8$. As usual in complex field theories, the Euler derivatives with respect to $e_\mu{}^a$ and its hermitian conjugate $e_{\mu a}$ are treated as independent. The symbol $\doteq$ denotes a field equation, distinguished from an identity. The symbol $I_\mu$ represents the imaginary part of $\Gamma_\mu \equiv \Gamma_{\mu\nu}{}^\nu$: $I_{\mu\nu}{}^\rho \equiv Im\,(\Gamma_{\mu\nu}{}^\rho)$, $I_\mu \equiv I_{\mu\nu}{}^\nu$.

Our field equation (3.9) can be confirmed based on the general invariance of our action (3.8) under

$$\delta_\xi e_\mu{}^a = \xi^\nu \partial_\nu e_\mu{}^a + (\partial_\mu \xi^\nu) e_\nu{}^a \ , \quad \delta_\xi e_a{}^\mu = \xi^\nu \partial_\nu e_a{}^\mu - (\partial_\nu \xi^\mu) e_a{}^\nu \ , \tag{3.11}$$

leading to the Noether identity[4]

$$\Big[ D_\mu F_\nu{}^\mu - (C_\mu - iI_\mu) F_\nu{}^\mu - C_{\nu\rho}{}^\tau F_\tau{}^\rho \Big] + \text{h.c.} \equiv 0 \ , \tag{3.12}$$

---
[4]As is well-known in conventional gravity theory [9] this identity is equivalent to a combination of Bianchi identity (2.5).

with the *real* parameter $\xi^\mu$. Here '+ h.c.' implies the addition of the hermitian conjugate of the preceding brackets. The covariant derivative $D_\mu$ in (3.12) contains only $\Gamma_{\mu\nu}{}^\rho$, while its hermitian conjugate $\overline{D}_\mu$ contains $\overline\Gamma_{\mu\nu}{}^\rho$. Note that eq. (3.12) is an *identity* based on the general invariance of our action, but *not* field equation.

Let us now consider the degrees of freedom of our field variables. There are originally 32 degrees of freedom for $e_\mu{}^a$ and $e_{\mu a}$, and 32 independent components in $\mathcal{F}^{\mu\nu}$. However, eventually we have 10 for $G_{\mu\nu}$ and 6 for $B_{\mu\nu}$ as the physical components. The gap between 32 and 16 should be understood as the redundancy of 16 components in $\mathcal{F}^{\mu\nu}$, by the use of

$$\frac{\delta e_\mu{}^a}{\delta G_{\rho\sigma}} = +\tfrac{1}{2}\delta_\mu{}^{(\rho|}e^{b|\sigma)}\eta_b{}^a \ , \qquad \frac{\delta e_a{}^\mu}{\delta G_{\rho\sigma}} = +\tfrac{1}{2}e_a{}^{(\rho|}g^{\mu|\sigma)} \ ,$$
$$\frac{\delta e_\mu{}^a}{\delta B_{\rho\sigma}} = -\tfrac{i}{2}\delta_\mu{}^{[\rho|}e^{b|\sigma]}\eta_b{}^a \ , \qquad \frac{\delta e_a{}^\mu}{\delta B_{\rho\sigma}} = +\tfrac{i}{2}e_a{}^{[\rho|}g^{\mu|\sigma]} \ , \tag{3.13}$$

leading to

$$\frac{\delta \mathcal{L}_0}{\delta G_{\mu\nu}} \equiv +Re\,(\mathcal{F}^{(\mu\nu)}) \ , \qquad \frac{\delta \mathcal{L}_0}{\delta B_{\mu\nu}} \equiv -Im\,(\mathcal{F}^{[\mu\nu]}) \ . \tag{3.14}$$

In other words, the components $Re\,(\mathcal{F}^{[\mu\nu]}) \doteq 0$ and $Im\,(\mathcal{F}^{(\mu\nu)}) \doteq 0$ yield extra constraints on the components in the vierbeins $e_\mu{}^a$ and $e_{\mu a}$ different from the directions of $G_{\mu\nu}$ and $B_{\mu\nu}$.

We next analyze our lagrangian (3.8) in terms of linearized gravity:

$$\begin{aligned}
e_\mu{}^a &= (\eta_\mu{}^a + h_\mu{}^a + a_\mu{}^a) + i(B_\mu{}^a + b_\mu{}^a) \equiv \eta_\mu{}^a + H_\mu{}^a \ , \\
e_{\mu a} &= (\delta_{\mu a} + h_{\mu a} + a_{\mu a}) - i(B_{\mu a} + b_{\mu a}) \equiv \delta_{\mu a} + \overline{H}_{\mu a} \ , \\
g_{\mu\nu} &= \delta_{\mu\nu} + h_{\mu\nu} + iB_{\mu\nu} \ , \qquad g^{\mu\nu} = \delta^{\mu\nu} - h^{\mu\nu} - iB^{\mu\nu} + \mathcal{O}(\varphi^2) \ , \\
h_{\mu a} &= +h_{a\mu} \ , \quad a_{\mu a} = -a_{a\mu} \ , \quad B_{\mu a} = -B_{a\mu} \ , \quad b_{\mu a} = +b_{a\mu} \ .
\end{aligned} \tag{3.15}$$

The fields $h_{\mu\nu}$, $a_{\mu\nu}$, $B_{\mu a}$, $b_{\mu a}$ are real, and $\mathcal{O}(\varphi^2)$ denotes any quadratic terms in fields. The $\delta_{\mu\nu}$ or $\delta^{\mu\nu}$ are the usual Minkowsky metric with the signature $(-,+,+,+)$, avoiding $\eta^{\mu\nu}$ which is confusing with $\eta_a{}^b$. The quadratic-order terms in (3.8) are now computed as

$$\begin{aligned}
\mathcal{L}_0|_{\text{quadratic}} =\ & +(+2a_1 - a_2)(\partial_\mu h_{\rho\sigma})^2 + (-2a_1 + a_2 + a_3)(\partial_\mu h^{\mu\nu})^2 \\
& - 2a_3(\partial_\mu h_\rho{}^\rho)(\partial_\nu h^{\mu\nu}) + a_3(\partial_\mu h_\rho{}^\rho)^2 \\
& + (+4a_1 - 2a_2 + 2a_3)\left[\,(\partial_\rho h_\mu{}^\rho)(\partial_\sigma a^{\mu\sigma}) + (\partial_\rho B_\mu{}^\rho)(\partial_\sigma b^{\mu\sigma})\,\right] \\
& + (+2a_1 + a_2)(\partial_\mu a_{\rho\sigma} + \partial_\mu B_{\rho\sigma})^2 + (-2a_1 - 3a_2 + a_3)(\partial_\nu a^{\mu\nu} + \partial_\nu B^{\mu\nu})^2 \\
& + (+2a_1 - a_2)(\partial_\mu b_{\rho\sigma})^2 + (-2a_1 + a_2 + a_3)(\partial_\nu b^{\mu\nu})^2 \\
& - 2a_3(\partial_\mu b_\rho{}^\rho)(\partial_\nu b^{\mu\nu}) + a_3(\partial_\mu b_\rho{}^\rho)^2 \ .
\end{aligned} \tag{3.16}$$

Since all the fields are real, we can use the usual Minkowsky metric $\delta_{\mu\nu}$ or $\delta^{\mu\nu}$ for contractions. The desirable aspects of our lagrangian are summarized as: (i) The kinetic terms for the antisymmetric fields $a_{\mu\nu}$ and $B_{\mu\nu}$, as well as all the $h$-$a$ and $B$-$b$ mixture terms disappear upon the condition (3.3); (ii) There is no mixture terms between the two sets $(h_{\mu\nu}, a_{\mu\nu})$ and $(B_{\mu\nu}, b_{\mu\nu})$, such as $aB$-terms, due to the hermiticity of the lagrangian; (iii) The kinetic terms for $h_{\mu\nu}$ coincide exactly with the quadratic terms of the Einstein-Hilbert action $S_{\text{EH}}$; (iv) At the quadratic level, there is no negative energy ghost, as $S_{\text{EH}}$ does not; (v) Interestingly, the field $b_{\mu\nu}$ also acquires its kinetic term, exactly with the same coefficients as that of $h_{\mu\nu}$, so that there are two sorts of spin 2 fields. We will come back to this point shortly.

We next analyze field equation (3.9) covariant to all orders. To this end, we need an additional constraint

$$Im\,(\Gamma_{\mu\nu}{}^{\rho}) \doteq 0 \quad . \tag{3.17}$$

Note that (3.17) is weaker than a direct condition $Im\,(g_{\mu\nu}) = 0$. However, at least by perturbation, we can show that the field equation (3.9) is satisfied with no bad effect on the propagation of the physical graviton $h_{\mu\nu}$.

We first note that the constraint (3.17) is covariant under general coordinate transformations (3.11):

$$\delta_\xi \Gamma_{\mu\nu}{}^{\rho} = \xi^\sigma \partial_\sigma \Gamma_{\mu\nu}{}^{\rho} + (\partial_\mu \xi^\sigma)\Gamma_{\sigma\nu}{}^{\rho} + (\partial_\nu \xi^\sigma)\Gamma_{\mu\sigma}{}^{\rho} - (\partial_\sigma \xi^\rho)\Gamma_{\mu\nu}{}^{\sigma} + \partial_\mu\partial_\nu \xi^\rho \quad . \tag{3.18}$$

With its hermitian conjugate subtracted, the last purely real term $\partial_\mu\partial_\nu\xi^\rho$ vanishes in the combination of $\delta_\xi[Im\,(\Gamma_{\mu\nu}{}^{\rho})]$, establishing its covariance. Now consider (3.17) in terms of the linearized gravity (3.15):

$$Im\,(\Gamma_{\mu\nu}{}^{\rho}) = (\partial_\mu B_\nu{}^\rho + \partial_\mu b_\nu{}^\rho) + \mathcal{O}(\varphi^2) = 0 \quad . \tag{3.19}$$

This condition is equivalent to two conditions

$$Im\,(\Gamma_{\mu[\nu\rho]}) = \partial_\mu B_{\nu\rho} + \mathcal{O}(\varphi^2) = 0 \quad , \quad Im\,(\Gamma_{\mu(\nu\rho)}) = \partial_\mu b_{\nu\rho} + \mathcal{O}(\varphi^2) = 0 \quad . \tag{3.20}$$

In other words, the components $B_{\mu\nu}$ and $b_{\mu\nu}$ are frozen with no space-time dependence. In fact, as long as Lorentz covariance is respected[5], the only solutions are $B_{\mu\nu} = \mathcal{O}(\varphi^2)$, $b_{\mu\nu} = \mathcal{O}(\varphi^2)$, so that $Im\,(e_\mu{}^a) = \mathcal{O}(\varphi^2)$, $Im\,(C_{\mu\nu}{}^a) = \mathcal{O}(\varphi^2)$. Therefore, we can conclude that at least perturbatively under the constraint $Im\,(\Gamma_{\mu\nu}{}^\rho) = 0$, all the linear terms in (3.9) are real up to cubic order terms: $Im\,(\mathcal{F}^{\mu\nu}) = \mathcal{O}(\varphi^3)$, and hence our field eq. (3.9) is equivalent to the general relativity up to the cubic terms, with no interference of extra components with the physical components $h_{\mu\nu}$. This nice feature is the result of our constraint (3.17), restricting the possible solutions for extra fields always one order higher: $B_{\mu\nu} = \mathcal{O}(\varphi^2)$, $b_{\mu\nu} = \mathcal{O}(\varphi^2)$. To put it differently, these extra fields do no harm on the propagation of the physical $h_{\mu\nu}$-field, up to cubic-order terms in the field equation, and quartic-order terms in the lagrangian.

---

[5]Lorentz covariance is respected, if we exclude solutions such as $B_{\mu\nu} = $ const. or $b_{\mu\nu} = $ const. For example, the particular case $B_{\mu\nu} = $ const. in noncommutative gravity is not included in this context for an obvious reason.

Analyzing the covariant field equation (3.9), we have seen that the additional spin 2 component $b_{\mu\nu}$ can be eliminated[6] by the constraint $Im(\Gamma_{\mu\nu}{}^{\rho}) \doteq 0$ imposed 'by hand', perturbatively up to the quartic terms in the lagrangian. Our next question is how to automatically implement such a constraint at the lagrangian level. To this end, we rely on the method in [10], using some multiplier field in a quadratic constraint lagrangian:

$$\mathcal{L}_{\Lambda} \equiv \tfrac{1}{2} |e| \Lambda^{\mu\nu\sigma\tau}{}_{\rho\lambda} Im(\Gamma_{\mu\nu}{}^{\rho}) Im(\Gamma_{\sigma\tau}{}^{\lambda}) \ . \tag{3.21}$$

The advantage of such constraint lagrangian [10] is not to affect the field equations of other non-constrained fields, due to the quadratic nature of this term.[7] In fact, the field equation for $\Lambda^{\mu\nu\sigma\tau}{}_{\rho\lambda}$ yields

$$Im(\Gamma_{\mu\nu}{}^{\rho}) Im(\Gamma_{\sigma\tau}{}^{\lambda}) \doteq 0 \tag{3.22}$$

for an arbitrary indices ${}_{\mu\nu\sigma\tau}{}^{\rho\lambda}$, if there is no index-symmetry for $\Lambda^{\mu\nu\sigma\tau}{}_{\rho\lambda}$. We can prove that (3.22) implies the constraint $Im(\Gamma_{\mu\nu}{}^{\rho}) \doteq 0$ by 'reduction to absurdity'. Since this proof is straightforward, we will skip it in this Letter.

After such a treatment, $h_{\mu\nu}$ in $G_{\mu\nu} = \delta_{\mu\nu} + h_{\mu\nu} = Re(g_{\mu\nu}) = g_{(\mu\nu)}$ is the only physical field with its kinetic terms in our formulation. In particular, the additional spin 2 component $b_{\mu\nu}$ has been deleted by our constraint lagrangian (3.21). Recall that this is due to the analysis from (3.19) and (3.20) leading to $B_{\mu\nu} = \mathcal{O}(\varphi^2)$, $b_{\mu\nu} = \mathcal{O}(\varphi^2)$.

We have another justification of our theory, based on the degrees of freedom. The extra components $a_{\mu\nu}$, $B_{\mu\nu}$ and $b_{\mu\nu}$ have respectively 6, 6 and 10 degrees of freedom. The field equations for $a_{\mu\nu}$, $B_{\mu\nu}$ and $b_{\mu\nu}$ are[8]

$$Re(\mathcal{F}^{[\mu\nu]}) \doteq 0 \ , \quad Im(\mathcal{F}^{[\mu\nu]}) \doteq 0 \ , \quad Im(\mathcal{F}^{(\mu\nu)}) \doteq 0 \ . \tag{3.23}$$

These independent components of the whole field equation $\mathcal{F}^{\mu\nu} \doteq 0$ give respectively $6+6+10$ equations to these unphysical auxiliary fields. On the other hand, we have seen from (3.20) that the constraint $Im(\Gamma_{\mu\nu}{}^{\rho}) \doteq 0$ is equivalent to freezing $6+10$ components $B_{\mu\nu}$ and $b_{\mu\nu}$. Even though the condition $Im(\Gamma_{\mu\nu}{}^{\rho}) = 0$ is weaker than $Im(g_{\mu\nu}) = 0$ due to possible cross-terms in $\Gamma_{\mu\nu}{}^{\rho}$, we have seen in (3.20) perturbatively that the extra components $B_{\mu\nu}$, $b_{\mu\nu}$ do no harm on the propagation of $h_{\mu\nu}$ in the field equation $Re(\mathcal{F}^{(\mu\nu)}) \doteq 0$. As for the possible bad effect by the remaining extra component $a_{\mu\nu}$, this has been already clarified based on the past teleparallel formulations [7]. Namely, once $B_{\mu\nu}$ and $b_{\mu\nu}$ become irrelevant, $Re(\mathcal{F}^{[\mu\nu]}) \doteq 0$ is automatically satisfied by Bianchi identities [7]. Therefore we conclude that there is perturbatively no undesirable effect on the propagation of $h_{\mu\nu}$ by the extra components $B_{\mu\nu}$ and $b_{\mu\nu}$ or $a_{\mu\nu}$.

---

[6]This additional spin 2 field may well have consistent interactions in our lagrangian (3.8), but we take rather a conservative viewpoint in this paper, leaving such a possibility for future studies.

[7]Even though there might be some subtlety with this constraint lagrangian at quantum level with path-integral, we do not go that far in this paper.

[8]In 'real' teleparallel gravity with the lagrangian (3.1), the equation $\mathcal{F}^{[\mu\nu]} \doteq 0$ is identically zero, due to the *hidden* local Lorentz invariance of the action (3.1). However, our action (3.8) has *no* such hidden local Lorentz invariance, as can be easily confirmed by the $U(1,3)$ local Lorentz analog of (3.6) with an additional derivative term $\partial_{\mu} \alpha_a{}^b$. Therefore, the second field equation in (3.23), which is equivalent to the $B_{\mu\nu}$-field equation *via* (3.6), is *not* identically zero.

Before summarizing this section, we analyze our field equation (3.9) for two important cases, when some of the extra components are absent:

(i) When the vierbein is real, the anholonomy coefficients becomes pure real $C_{\mu\nu}{}^\rho = \overline{C}_{\mu\nu}{}^\rho$, and $I_\mu = 0$, while the covariant derivative $D_\mu$ coincides with its hermitian conjugate $\overline{D}_\mu$, and in particular, $D_\mu g_{\nu\rho} \equiv 0$:

$$|e|^{-1}\mathcal{F}^{\mu\nu}\Big|_{\mathfrak{Im}(g_{\rho\sigma})=0} = -4aD_\rho C^{\rho[\mu\nu]} + 4ag^{\mu\nu}D_\rho C^\rho - 4aD^\mu C^\nu - 2aD_\rho C^{\mu\nu\rho}$$
$$- 4aC_\rho C^{\rho(\mu\nu)} - 2aC^{\mu\rho}{}_\sigma C^\nu{}_\rho{}^\sigma + aC_{\rho\sigma}{}^\mu C^{\rho\sigma\nu}$$
$$- 2aC^\mu{}_\rho{}^\sigma C^\nu{}_\sigma{}^\rho - 2aC_\rho C^{\mu\nu\rho} + 2ag^{\mu\nu}(C_\rho)^2 + \tfrac{1}{2}ag^{\mu\nu}(C_{\rho\sigma\tau})^2 \quad . \quad (3.24)$$

The symmetric terms $\mathcal{F}^{(\mu\nu)}$ can be shown to be equivalent to the usual Einstein tensor in general relativity in terms of anholonomy coefficients [7]. On the other hand, all the antisymmetric terms $\mathcal{F}^{[\mu\nu]}$ cancel each other upon the Bianchi identity (2.5). We can also mimic this procedure at the lagrangian level (3.8), reducing (3.8) into the Einstein-Hilbert action (3.1). Note that we have *not* imposed the absence of the antisymmetric part $a_{\mu\nu}$, but it is decoupled from the whole field equation. There is nothing unusual about this, because upon the constraint $Im(e_\mu{}^a) = 0$, our system is reduced to the teleparallel 're-formulation' of general relativity with no kinetic term for $a_{\mu\nu}$ as in [7], and thus it is equivalent to the ordinary general relativity.

(ii) When the vierbein and metric are complex constant, then $\Gamma_{\mu\nu}{}^\rho \equiv 0$ holds due to the postulate (2.7), leading to $C_{\mu\nu}{}^\rho \equiv 0$. Accordingly, the covariant derivatives $D_\mu = \overline{D}_\mu$ become the ordinary derivative $\partial_\mu$. Therefore, each term in (3.9) vanishes identically, confirming that such a metric is a trivial solution to our field equation. This aspect is important, since we can now have the constant but complex metric $g^{\mu\nu} = \delta^{\mu\nu} + i\theta^{\mu\nu}$ used in noncommutative gravity [1].

To summarize, we have obtained the following important results:

(1) The analysis of the quadratic terms in our lagrangian (3.8) shows no negative energy ghosts. In terms of linearized fields, $h_{\mu\nu}$ is the ordinary graviton, while $a_{\mu\nu}$ and $B_{\mu\nu}$ are auxiliary with no kinetic terms. In particular, the $B_{\mu\nu}$-field has no negative energy ghost, in contrast to [2]. There is an additional spin 2 field $b_{\mu\nu}$ with physical kinetic terms in (3.16), which can be further deleted by a constraint lagrangian (3.21), as in items (4) and (5) below.

(2) The field equations $Re(\mathcal{F}^{[\mu\nu]}) \doteq 0$, $Im(\mathcal{F}^{[\mu\nu]}) \doteq 0$ and $Im(\mathcal{F}^{(\mu\nu)}) \doteq 0$ provide $6 + 6 + 10 = 22$ equations. These are $6 + 6$ constraints for the extra non-propagating components $a_{\mu\nu}$, $B_{\mu\nu}$ and 10 field equations for spin 2 field $b_{\mu\nu}$. Moreover, the constraint $Im(\Gamma_{\mu\nu}{}^\rho) = 0$ can freeze $B_{\mu\nu}$ and $b_{\mu\nu}$, as in item (4) below. This constraint is also covariant under general coordinate transformation.

(3) The constraint $Im(\Gamma_{\mu\nu}{}^\rho) = 0$ is automatically realized, by adding the constraint lagrangian $\mathcal{L}_\Lambda$ (3.21), as in [10].

(4) Under the constraint $Im\,(\Gamma_{\mu\nu}{}^\rho) \doteq 0$, the only perturbative solutions are $B_{\mu\nu} = \mathcal{O}(\varphi^2)$ and $b_{\mu\nu} = \mathcal{O}(\varphi^2)$. Hence all the worrisome quadratic terms with $B_{\mu\nu}$, $b_{\mu\nu}$ in the field equation $Re\,(\mathcal{F}^{(\mu\nu)}) \doteq 0$ of $h_{\mu\nu}$ become actually cubic order (one-order higher). Therefore, at least perturbatively, they will do no harm on $h_{\mu\nu}$ at the quadratic order at the field equation level, and the cubic order at the lagrangian level.

(5) Our covariant field equation (3.9) is reduced to the standard Einstein equation in general relativity, when the vierbeins becomes real. In other words, any real vierbein solution to the Einstein equation automatically satisfies our field equation as a special case. This corresponds to the symmetric real part $Re\,(\mathcal{F}^{(\mu\nu)}) \doteq 0$.

(6) The constant but generally complex metric, breaking the ordinary Lorentz covariance, is a simple solution to our field equation (3.9). This includes the important case of $g^{\mu\nu} = \delta^{\mu\nu} + i\theta^{\mu\nu}$.

Before closing this section, we mention one possibility of generalizing our lagrangian (3.8). Based on teleparallel real gravity in [7], we can add the kinetic term

$$\mathcal{L}_4 \equiv a_4 |e|^{-1} g_{\tau\mu} \epsilon^{\mu\nu\rho\sigma} \epsilon^{\tau\lambda\omega\psi} C_{\nu\rho\sigma} \overline{C}_{\lambda\omega\psi} \quad, \tag{3.25}$$

so that the component $a_{\mu\nu}$ is now propagating with its kinetic term like $(\partial_{[\mu} a_{\nu\rho]})^2$ with no negative energy states, with an appropriate sign for $a_4$. As long as we have *no* term like (3.25), all the terms in our lagrangian (3.8), apart from those produced by the imaginary part of $e_\mu{}^a$, coincide with the Einstein-Hilbert lagrangian, even if there are antisymmetric components $a_{\mu\nu}$ in $e_\mu{}^a$ [7]. This is due to a 'hidden' local Lorentz invariance of our action when $Im\,(e_\mu{}^a) = 0$. This feature has been well-known in teleparallelism formulation [7].

## 4. Key Classical Test of Perihelion Advance

We have seen that our teleparallel complex gravity has no negative energy ghosts based on the quadratic order analysis. Moreover, our system reproduces general relativity at the cubic order in the lagrangian with no bad effect by the extra components. However, some readers may be wondering if our system passes the important key classical tests with more higher-order interactions, like the perihelion advance of Mercury.

To study this, we start by postulating an action for a point mass:

$$I_\mathrm{M} = \int ds \left( G_{\mu\nu} \frac{dx^\mu}{ds} \frac{dx^\nu}{ds} \right)^{1/2} \quad. \tag{4.1}$$

Here we have put only the real part of our metric $G_{\mu\nu} \equiv Re\,(g_{\mu\nu}) = G_{\nu\mu}$. Even though it is natural to have only the real part $G_{\mu\nu}$ in $I_\mathrm{M}$, since we do not have the total metric $g_{\mu\nu}$ here, the general coordinate invariance $\delta_\xi I_\mathrm{M} = 0$ under (3.11) is non-trivial. However, this can be easily confirmed, once we realize from (3.11) that

$$\delta_\xi G_{\mu\nu} = \xi^\rho \partial_\rho G_{\mu\nu} + (\partial_\mu \xi^\rho) G_{\rho\nu} + (\partial_\nu \xi^\rho) G_{\mu\rho} \quad. \tag{4.2}$$

In other words, the lack of the imaginary part $Im(g_{\mu\nu})$ does not upset the desirable transformation property of $G_{\mu\nu}$. Due to this property (4.2), it is now clear that our action $I_M$ has the general coordinate invariance $\delta_\xi I_M = 0$.

Once we have put only the real part $G_{\mu\nu}$ into $I_M$, the usual geodesic equation of motion for a point mass follows, exactly as in general relativity:

$$\frac{d^2 x^\mu}{ds^2} + \left\{ {}^{\mu}_{\rho\sigma} \right\} \frac{dx^\rho}{ds} \frac{dx^\sigma}{ds} \doteq 0 \quad , \tag{4.3}$$

with $\left\{ {}^{\mu}_{\rho\sigma} \right\} \equiv (1/2) G^{\mu\tau} (\partial_\rho G_{\sigma\tau} + \partial_\sigma G_{\rho\tau} - \partial_\tau G_{\rho\sigma})$ involving only $G_{\mu\nu}$ and its inverse $G^{\mu\nu}$.

Our next question is whether the conventional Schwarzschild metric

$$ds^2 = -\left(1 - \frac{2m}{r}\right) + \frac{1}{1 - \frac{2m}{r}} dr^2 + r^2 (d\theta^2 + \sin^2\theta \, d\varphi^2) \tag{4.4}$$

satisfies our gravitational field equation (3.9). However, as explained in item (5) at the end of the last section, this Schwarzschild metric is indeed a solution to (3.9). In fact, since $F^{\mu\nu} \doteq 0$ is reduced to the conventional Einstein tensor $R^{\mu\nu}(G) - (1/2) G_{\mu\nu} R \doteq 0$ upon the restriction $Im(g_{\mu\nu}) = 0$ [7], the Schwarzschild metric (4.4) indeed satisfies our gravitational field equation (3.9), as the simplified case. On the other hand, we have seen that a point mass satisfies exactly the same geodesic equation (4.2) as in general relativity in terms of $G_{\mu\nu}$. Therefore, the computation of perihelion advance in our theory exactly agrees with that in general relativity.

We mention also the compatibility of our theory with equivalence principle. Since the experimental tests of 'equivalence principle' so far have been always for classical macroscopic objects, it is enough to consider only the point mass action $I_M$. As we have seen, it is only the real Christoffel connection $\left\{ {}^{\rho}_{\mu\nu} \right\}$ that couples to a point mass. If that is the case, all the extra components $a_{\mu\nu}$, $B_{\mu\nu}$ and $b_{\mu\nu}$ are irrelevant to the question of equivalence principle associated with point masses.[9]

To put it differently, we have seen that our constraint $Im(\Gamma_{\mu\nu}{}^\rho) \doteq 0$ (3.17) deletes the covariant and imaginary part of the affinity $\Gamma_{\mu\nu}{}^\rho$, while the real part $Re(\Gamma_{\mu\nu}{}^\rho)$ obeys the same transformation rule as general relativity. Therefore, by choosing a geodesic coordinate system under (3.17), we can make also the real part $Re(\Gamma_{\mu\nu}{}^\rho)$ vanish, as in general relativity. Eventually, the geodesic equation (4.3) becomes a 'free-fall' equation $d^2 \widetilde{x}^\mu / ds^2 \doteq 0$, as in general relativity. As long as we use a classical macroscopic object as a test point mass, we can not tell the violation of equivalence principle in our theory.

We emphasize that our gravitational theory has indeed passed the most important key test of perihelion advance of Mercury and equivalence principle, at least at the classical field level.

---

[9]As for other questions related to other particle theoretic fields with different spins such as fermions, we do not discuss them here, because they are beyond the scope of this Letter.

## 5. A Lagrangian for Noncommutative Gravity

Once we have understood the teleparallelism formulation of complex gravity, it is straightforward to generalize that result to the noncommutative gravity, with the standard $\star$ product:

$$f \star g \equiv f \exp{(i\overleftarrow{\partial}_\mu \theta^{\mu\nu} \overrightarrow{\partial}_\nu)} g \equiv \sum_{n=1}^{\infty} \frac{i^n}{n!} \theta^{\mu_1\nu_1} \cdots \theta^{\mu_n\nu_n} (\partial_{\mu_1} \cdots \partial_{\mu_n} f)(\partial_{\nu_1} \cdots \partial_{\nu_n} g) \ . \quad (5.1)$$

The hermiticity of a noncommutative product of complex functions $f_1, \cdots, f_n$ is understood as $(f_1 \star f_2 \star \cdots \star f_{n-1} \star f_n)^\dagger = f_n^\dagger \star f_{n-1}^\dagger \star \cdots \star f_2^\dagger \star f_1^\dagger$. We can replace all the products and matrix inverses by $\star$ products and $\star$ inverses everywhere in (3.8) + (3.21):

$$\begin{aligned} \mathcal{L}_\star \equiv &+ \sqrt[4]{e_\star}^\dagger \star \Big[ -\tfrac{1}{8} g_\star^{\mu\rho} \star C_{\mu\nu}{}^a \star \eta_a{}^b \star \overline{C}_{\rho\sigma b} \star g_\star^{\nu\sigma} + \tfrac{1}{4} e_\star^{b\nu} \star C_{\mu\nu}{}^a \star g_\star^{\mu\sigma} \star \overline{C}_{\rho\sigma b} \star e_{\star a}{}^\rho \\ &+ \tfrac{1}{2} e_{\star a}{}^\nu \star C_{\mu\nu}{}^a \star g_\star^{\mu\rho} \star \overline{C}_{\rho\sigma b} \star e_\star^{b\sigma} \Big] \star \sqrt[4]{e_\star} \\ &+ \tfrac{1}{2} \sqrt[4]{e_\star}^\dagger \star Im\,(\Gamma_{\mu\nu}{}^\rho) \star \Lambda^{\mu\nu\sigma\tau}{}_{\rho\lambda} \star Im\,(\Gamma_{\sigma\tau}{}^\lambda) \star \sqrt[4]{e_\star} \ , \end{aligned} \quad (5.2)$$

paying attention to comply with the hermiticity mentioned above. Here $e_{\star a}{}^\mu \star e_\mu{}^b = \delta_a{}^b$ and $e_\star^{b\nu} \equiv (e_{\star b}{}^\nu)^\dagger$, etc., similarly to [2]. Due to the $\star$ product, we need the subscript $\star$ for the inverse vierbeins, such as $e_{\star a}{}^\mu$ and $e_\star^{b\nu}$. Relevantly, $e_\star$ is the $\star$-determinant of the vierbein $e_\mu{}^a$: $e_\star \equiv (1/4!)\epsilon_{abcd}\epsilon^{\mu\nu\rho\sigma} e_\mu{}^a \star e_\nu{}^b \star e_\rho{}^c \star e_\rho{}^d$. The operation $\sqrt[4]{e_\star}$ can be understood as the expansion $\sqrt[4]{1+x} \equiv 1 + \sum_{n=1}^{\infty}(1/n!)(1/2)(1/2-1)\cdots(3/2-n) \overbrace{x \star x \star \cdots \star x}^{n}$ [11], while $\sqrt[4]{e_\star}^\dagger$ is its hermitian conjugate. Since $g_{\mu\nu}$ is hermitian, there is no problem with the product for the definition of the square roots in $\sqrt[4]{e_\star}^\dagger \star \sqrt[4]{e_\star}$ as in the commutative case. The definition of $C_{\mu\nu}{}^a$ or $\overline{C}_{\mu\nu a}$ needs no $\star$-symbol on itself like $C_{\star\mu\nu}{}^a$, because it has no product in its definition just as in the commutative case (3.8). Thanks to teleparallel complex gravity as the foundation, we no longer have the problem with the negative energy states in our system, while all the unphysical components are either decoupled, or do no harm to the physical components $G_{\mu\nu} \equiv Re\,(g_{\mu\nu}) \equiv g_{(\mu\nu)}$. As for the uniqueness of our lagrangian (5.2), we are aware that there might be other possible forms of lagrangians depending on the order of the terms in (5.2). For example, the metrics and the anholonomy coefficients might be flipped around in the first term in (5.2), as long as the hermiticity is satisfied. However, we take here the standpoint that any of these lagrangians shares the same quadratic terms in the teleparallel gravity, and we gave (5.2) as an explicit example, whose quadratic terms are shared by any of those other examples.

## 6. Concluding Remarks

In this Letter, we have presented teleparallel complex gravity as the starting point for noncommutative gravity. Since the introduction of the $\star$ product implies that the metric is to be complex, and the problem of the negative energy ghost is caused by the Lorentz connection [2], it is natural to consider the formulation without manifest $U(1,3)$ local

Lorentz symmetry equivalent to teleparallelism, motivating the combination of teleparallel complex gravity.

We have studied the transformation properties under general coordinate transformations and global Lorentz transformations. We have next presented a lagrangian (3.8) that contains not only the usual Einstein-Hilbert action for the symmetric part $G_{\mu\nu} \equiv g_{(\mu\nu)} \equiv Re\,(g_{\mu\nu})$ but also the imaginary part $B_{\mu\nu} \equiv Im\,(g_{\mu\nu}) = -ig_{[\mu\nu]}$. Our lagrangian (3.8) has an improved property that there is no kinetic term with negative energy states for $B_{\mu\nu} \equiv -ig_{[\mu\nu]}$. This has been proven explicitly with the lagrangian quadratic terms (3.16), which show that all the kinetic terms for $a_{\mu\nu}$ and $B_{\mu\nu}$ disappear under the condition (3.3), while the component $b_{\mu\nu}$ has a kinetic term as a spin 2 field with positive definite energy. Thus there is no negative energy ghost in our system. We have next looked into the field equation $\mathcal{F}^{\mu\nu} \doteq 0$ of the vierbein $e_\mu{}^a$, and concluded that the real symmetric part $Re\,(\mathcal{F}^{(\mu\nu)}) \doteq 0$ contains the Einstein gravitational field equation for the symmetric component $g_{(\mu\nu)}$, while other remaining parts $Re\,(\mathcal{F}^{[\mu\nu]}) \doteq 0$, $Im\,(\mathcal{F}^{[\mu\nu]}) \doteq 0$ and $Im\,(\mathcal{F}^{(\mu\nu)}) \doteq 0$ yield $6+6+10 = 22$ component general covariant constraint equations respectively for the extra $6+6+10 = 22$ auxiliary non-propagating components $a_{\mu\nu}$, $B_{\mu\nu}$ and $b_{\mu\nu}$, as desired. We have also seen that the Lorentz non-covariant solution $\theta^{\mu\nu} \equiv Im\,(g^{\mu\nu}) = $ const. is also an acceptable solution, which is important for the noncommutative gravity [1].

We have also confirmed that the covariant constraint $Im\,(\Gamma_{\mu\nu}{}^\rho) = 0$ is enough to freeze the extra components $B_{\mu\nu}$, $b_{\mu\nu}$ in such a way that the physical component $h_{\mu\nu}$ is not constrained for its propagation. This is due to the only perturbatively possible solutions $B_{\mu\nu} = \mathcal{O}(\varphi^2)$, $b_{\mu\nu} = \mathcal{O}(\varphi^2)$ for the constraint $Im\,(\Gamma_{\mu\nu}{}^\rho) \doteq 0$, which are one order higher in terms of fields. Based on this nice feature, we have confirmed no disturbing effect on the physical components $G_{\mu\nu}$ from these extra components up to cubic-order terms in the field equation, equivalent to quartic-order terms in the lagrangian. We emphasize that our total lagrangian (3.8) + (3.21) 'reproduces' general relativity to this order, with no bad effect from the extra components.

As an important test of classical gravitational theory, we have analyzed the possible effect of the extra components on the perihelion advance. Based on our point mass action $I_M$, we have concluded that there is no disturbing effect on the perihelion advance, due to the absence of the imaginary part of the metric $Im\,(g_{\mu\nu})$ in $I_M$. Our result has excluded a worry that some extra components of our teleparallel complex gravity might interfere with such a high precision observation as the perihelion advance of Mercury.

Based on this result of teleparallel complex gravity, we have presented a lagrangian for noncommutative gravity with no negative energy ghosts. Even though computations for teleparallelism formulation seem to be more involved compared with [2], the advantage here is that the problem with the negative energy states in $B_{\mu\nu}$-field in [2] has been now resolved.

The main purpose of our present paper is to establish teleparallel complex gravity as the foundation for noncommutative gravity. However, our result also suggests the importance of teleparallel gravity even *before* its complexification in string physics, M-theory or D-branes. This further indicates teleparallel supergravity playing an important role in superstring or M-theory. As a matter of fact, a teleparallel superspace had already been considered in 1970's in four-dimensions (4D) as a possible resolution to renormalizability problem [12], by reducing the number of counter-terms in the absence of supercurvature. Also in our recent paper [13], we have constructed a 11D teleparallel superspace as a reformulation of 11D supergravity, not only as a consistent but also as a more natural background for supermembranes. From these developments, we regard teleparallel supergravity or teleparallel superspace as a more natural formulation suitable for the description of (super)strings, M-theory or D-branes.

Considering the fundamental fact that the constant tensor $\theta^{\mu\nu}$ in noncommutative gravity manifestly breaks Lorentz symmetry, we also emphasize that teleparallelism is the most natural candidate for the foundation of noncommutative gravity.

We are grateful to A. Chamseddine and S.J. Gates, Jr. for helpful discussions. A special acknowledgment is due to the referee(s) of this paper who suggested to analyze the quadratic lagrangian terms and the perihelion advance of Mercury.

# References

[1] A. Connes, M.R. Douglas and A. Schwarz, JHEP **9802** (1998) 003; Y.K.E. Cheung and M. Krogh, Nucl. Phys. **B528** (1998) 185; C.-S.Chu and P.-M. Ho, Nucl. Phys. **B528** (1999) 151; V. Schomerus, JHEP **9906** (1999) 030; F. Ardalan, H. Arfaei and M.M. Sheikh-Jabbari, JHEP **9902** (1999) 016; J. Hoppe, Phys. Lett. **B250** (1990) 44; D.B. Fairlie, P. Fletcher and C.K. Zachos, Phys. Lett. **B218** (1989) 203; N. Seiberg and E. Witten, JHEP **9909** (1999) 032, hep-th/9908142.

[2] A. Chamseddine, Commun. Math. Phys. **218** (2001) 283, hep-th/0005222; Int. Jour. of Mod. Phys. **A16** (2001) 759, hep-th/0010268.

[3] T. Damour, S. Deser and J. McCarthy, Phys. Rev.**D47** (1993) 1541.

[4] L. Bonora, M. Schnabl, M.M. Shaikh-Jabbari and A. Tormasiello, Nucl. Phys. **B589** (2000) 461, hep-th/0006091; I. Bars, M.M. Sheikh-Jabbari and M.A. Vasiliev, Phys. Rev. **D64** (2001) 086004, hep-th/0103209.

[5] A. Chamseddine, Phys. Lett. **504B** (2001) 33.

[6] J.W. Moffat, Phys. Lett. **B491** (2000) 345, hep-th/0007181.

[7] K. Hayashi and T. Nakano, Progr. Theor. Phys. **38** (1967) 491; Y.M. Cho, Phys. Rev. **D14** (1976) 2521, 3335 and 3341; F.W. Hehl, P. von der Heyde, D. Kerlick and J.M. Nester, Rev. Mod. Phys. **48** (1976) 393; K. Hayashi and T. Shirafuji, Phys. Rev. **D19** (1979) 3524.

[8] E. Langmann and R.J. Szabo, Phys. Rev. **D64** (2001) 104019, hep-th/0105094.

[9] *See, e.g.,* C.W. Misner, K.S. Thorne and J.A. Wheeler, *'Gravitation'* (W.H. Freeman and Company, NY, 1972); S. Weinberg, *'Gravitation and Cosmology'* (John Wiley & Sons, NY, 1971); L.D. Landau and E.M. Lifshitz, *'The Classical Theory of Fields'*, 4th English ed., Pergamon Press (1975).

[10] W. Siegel, Nucl. Phys. **B238** (1984) 307.

[11] *See, e.g.,* N. Grandi, R.L. Pakman and F.A. Schaposnik, Nucl. Phys. **B588** (2000) 508, hep-th/0004104.

[12] J.G. Taylor, Phys. Lett. **78B** (1978) 577.

[13] S.J. Gates, Jr., H. Nishino and S. Rajpoot, Phys. Rev. **D65** (2002) 024013, hep-th/0107155.